\documentclass[12pt]{amsart}
\usepackage{geometry} 
\usepackage{graphicx}

\title{Galactic Spiral Arms: Structure and Dynamics Given by an Equation of Motion }
\author{Gregory Buck}
\date{March 1, 2013}

\begin{document}
\maketitle 

{\bf Using an equation of motion for a self-gravitating filament, we show how galactic spiral arms might be created and sustained. We find that the combination of differential rotation of the galactic disk and the self-gravity of the arm (as given by the equation) leads to a rotating spiral structure. Moreover, using this analysis, we then find a second differential equation that explicitly relates this spiral structure to the rotation curve of the galaxy -- it connects several factors, including spiral shape and pattern speed.  We also describe a simple many-body numerical experiment that supports our approach. The findings are with consistent with observational evidence concerning arm structure and rotation curves, including leading arm structures.  }

\medskip

One school of thought holds that spiral arms of galaxies cannot be material -- rotation curves imply that material arms, that is, arms populated with the same stars and particles, would wrap up as the galaxy turns [1,2].  Therefore it seems that the spiral arm structure must be some kind of density wave -- the stars rotate at the speeds given by the rotation curve, but somehow cluster along the spiral arms in the process.  Various mechanisms have been offered for the dynamics of arms [1,3]. Others believe that the arms are more transient in nature -- breaking up before the differential rotation has had a chance to fully wrap them up [4,5,6,7]. Some think that a central bar is an important factor in the formation of the arms [8], others have put forward the notion that the arms are the result of interactions with another galaxy [9].  In any case, the  formation and sustenance of spiral arms might fairly be called a topic at least open for discussion.  Here we offer a simple mechanism that can support either transient or long-term arms.

In [10] is introduced an equation of motion for a self-interacting filament, it applies in the gravitational case [11].  It is $${\ddot \alpha =}C [2 \rho^\prime T + \rho \kappa N],$$
where  $\alpha(s)$  is the filament parametrized by arclength $s$,  $\rho(s)$ is the mass density along $\alpha$, $\rho^\prime(s)$ the rate of change of the density with respect to $s$, $T(s)$ the unit tangent vector, $N(s)$ the unit normal vector and $\kappa(s)$ the curvature along $\alpha(s)$, $C$ is a constant, and
a dot signifies the time derivative of the motion.

The derivation of this equation is akin to the Biot-Savart cut-off approach to the equation of motion for a line vortex.  The idea is that since a one-dimensional distribution of mass (a mass density on a space curve) is a divergent object under self-gravity, we can analyze the singularity to arrive at the equation.  This is a local theory, in that it does not account for mass that is close to the test point but which is not on the curve, nor for mass that is distal. 

We apply this equation to a spiral arm, and for the moment focus on the $\kappa$ term -- for momentary illustrative simplicity assume that $\rho^\prime =0$ along the arm.   (One could assume the arm mass density is given by the exponential decline with radius usually assumed for disk galaxies -- we do in models below -- and note then that since the spiral arm goes around the center as well as away, its exponential decline in mass is slower than along a ray of the disk, so $\rho^\prime$ is very small).    

A spiral has decreasing curvature as we follow the arc away from the center.  If we assume a mass distribution along a spiral, then by the equation of motion we have, due to the self-gravitation of the arm, an acceleration at each point in the normal direction proportional to the (decreasing) curvature.  For a trailing arm spiral, these vectors point "in" (have a component toward the center), and backward -- against the direction of rotation.  

A typical flat or decreasing rotation curve implies shear -- differential rotation.  In order for the spiral pattern to be sustained, each point in the arm must rotate with a speed roughly proportional to its distance from the center, for some constant proportion over the entire arm.  The mechanism we propose is that the self-gravitation directs the masses in and backwards, more strongly near the center and diminishing as we move away, and this compensates for the shear, allowing for a rigidly rotating pattern.  So, if we are thinking of the arm as a density wave (as opposed to material) the orbits of the stars can be thought of as roughly circular with speed given by the rotation curve for motion outside the arm.  As they enter the spiral arm, they are under the additional influence of the arm, directed backward and toward the center.  Their rotational speed is diminished, creating the density wave that is the spiral arm.  They then exit the arm to continue around the center. See figures 1 and 2. 

This mechanism is consistent with several aspects of spiral arms.  Since the argument is largely local, it can support any number of arms, including so-called flocculent structures, where there are many vaguely defined arms.  There is evidence of a correlation between spiral pitch angle and shear rate -- greater shear tends to give a lesser pitch angle [6,7,8].  This is natural in our mechanism -- the lesser pitch angle turns the normal vector more towards the center, allowing the resonance to be accomplished with less force from the arm.  See figure 2.

There is also evidence of leading (as opposed to trailing) spiral arms -- these are arms that open into to direction of rotation [9,10].  We can find a similar resonance pattern for this sort of arm, though in this case we require a rotation curve such that outside layers move past inside ones in the direction of rotation.   The observational evidence is that leading arms have such rotation curves.  In this case our mechanism gives a spiral pattern that rotates faster at each radial distance than the speed given by the rotation curve.  See figure 2.

Under this mechanism, the spiral arm is a stable structure -- an attractor.  If the self-gravity of the arm, as given by the curvature, is not great enough, then the arm winds up further, thus creating greater curvature.  On the other hand, if the curvature is too great, it defeats the winding, flattening the arm -- decreasing curvature.

We can make a simplifying assumption that allows us to mathematically model spiral arm structure and motion in two ways.  We run an elementary numerical  many-body experiment, and we find a differential equation, which depends on the rotation curve of the galaxy and the pattern speed, the solution of which is the spiral arm (we solve it numerically).  The simplifying assumption is that the masses keep (approximately) their radius with respect to the center of the galaxy, and so the gravitational forces of the arm serve to "locally" speed up or slow them down along these roughly  circular orbits. We also offer a caveat.  In the equation we are combining two different sorts of quantities: the gravitational force of the arm as given by the filament equation of motion, and rotational velocities. (Of course combining forces is natural, as is combining velocities, but not necessarily a mix of the two).  We argue that because of the force from the arm, a mass passing through the arm spends more time in the vicinity of the arm location than it would otherwise, thus changing the rotational velocity, and the magnitude of this change is proportional to the force. Some justification for this view can be found in figure 3, where we have numerically found the trajectory for a mass approaching and passing through a filamentary arm.  We can see clearly that the amount of time spent near the arm depends on the curvature of the arm.  We also note that some models treat the masses in the galactic disk as if there is a kind of friction among them, that collisions and near passes and gas and particle clouds tend to keep masses at the same radius moving at the same speed.  This tendency can be thought of as a force, pushing velocities toward certain values.  If viewed this way, we can think of our argument as a sort of balance of forces between this 'force' and the gravitational force of the arm. 

The numerical experiment is very simple.  We start a group of masses in some sort of roughly filamentary distribution, with one end at the center.  We then apply differential rotation to this distribution, moving the masses in circular orbits, according to the rotation speed associated with their radius.  We observe that the arm is stretched by this, and eventually the masses are so dispersed that there is no evidence of the original filamentary structure.  We then run the experiment again, this time adding a discretized version of the curvature force (a local action between nearest neighbors), at each time step in the experiment.  In this case the result is that we see persistent spiral structures.  For some parameter values these structures appear permanent, rotating for as long as the computer runs, for others they seem be in the constant process of making and unmaking, finishing several revolutions before breaking up then reforming. See figure 4.

The equation with solutions giving shape and motion of the spiral is the following:

$$ V(r)+A \| Proj_{rot} ( \rho \kappa N)\| +B \|Proj_{rot} (2\rho' T ) \| =\Omega_p r$$

Here $r$ is the radius from the center of the galaxy. $V(r)$ is the rotation curve for the galaxy. $\Omega_p $ is the constant pattern speed -- assuming the arm structure rotates rigidly about the center of the galaxy.  If a changing pattern is desired, we can replace $\Omega_p r$  with another function of $r$, $\Omega_p (r)$ .    $\kappa $ is the curvature, $\rho$  the effective density at a given station along the arm, $\rho'$ the rate of change of the density along the arm at a given station, $Proj_{rot}$ means that we project this vector onto the tangent of the rotation at the given station (recall that our simplifying assumption is that  the gravitational force of the arm increases or decreases the rotational speed).   $A$, $B$ are constants, which may be required because the velocities and the density have differing units, curvature depends on scale, etc. The sign of $A$ depends on whether or not the arm curve opens toward the direction of rotation -- for example for a trailing arm $A<0$ if the motion given by $V(r)$ is in the counterclockwise (positive) direction, so the force from this term is in the clockwise (negative) direction.  The sign of $B$ depends on whether the density increases or decreases along the curve in the direction of the rotation.  For a typical trailing arm spiral we have $B>0$.

If we can write all quantities as functions of radius $r$, then our equation becomes an ordinary differential equation with independent variable $\theta$ and dependent variable $r$, so with polar solution $r(\theta)$.

 Let $\beta $ be the angle of inclination of the arm at a given station (a circle about the center of the galaxy has $\beta=0$, a ray from the center of the galaxy has $\beta ={ \pi  \over 2}$). Then $sin ( \beta  )$, $cos( \beta  )$  will project the force, as given by  $\rho \kappa N$ and $2 \rho' T$, onto the tangent of the rotation.  But $\beta (r)  = arctan({r'( \theta) \over r (\theta)})$, and $\kappa (r) ={ |r^2 + 2r'^2 -r r''| \over (r^2 + r'^2)^{3\over 2}}$.  One way of approaching the density is to let the density of the arm simply be the density of the disk, that is, we assume that the density of the disk is given by a radially symmetric function, so $\rho = \rho_{disk} (r)$; this function is usually assumed to be some sort of exponential: $\rho_{disk} (r) = c_1 e^{c_2 r}$ for the portion of the disk not in the galactic bulge near the center.  Whatever the function is, we would have $\rho' (r) = \rho_{disk}'(r) sin(\beta)$, because the rate of change of the density along the curve would depend on the angle of inclination.

Then the equation becomes:

$$V(r)+ Asin \left( arctan{\left[ {r' \over r }\right]}\right) \rho(r) { |r^2 + 2r'^2 -r r''| \over (r^2 + r'^2)^{3\over 2}}+2Bcos\left( arctan{\left[ {r' \over r }\right]}\right) \rho' (r)=\Omega_p r$$

For an exponential disk this is:

$$V(r)+ Asin \left( arctan{\left[ {r' \over r }\right]}\right) c_1 e^{c_2 r}{ |r^2 + 2r'^2 -r r''| \over (r^2 + r'^2)^{3\over 2}}$$ $$+2Bcos\left( arctan{\left[ {r' \over r }\right]}\right) c_1c_2 e^{c_2 r}sin \left( arctan{\left[ {r' \over r }\right]}\right)=\Omega_p r$$

We also observe that an arm with constant density has $\rho' =0$ (this might be a reasonable approximation if the exponential rate $c_2$ is small enough), so drops that term and we have:

$$V(r)+ Asin \left( arctan{\left[ {r' \over r }\right]}\right) { |r^2 + 2r'^2 -r r''| \over (r^2 + r'^2)^{3\over 2}}=\Omega_p r$$

or

$$V(r)+{A r' \over  r {\sqrt {1+ ({r' \over r })^2}}} { |r^2 + 2r'^2 -r r''| \over (r^2 + r'^2)^{3\over 2}}=\Omega_p r$$

$V(r)$ has been estimated for many galaxies, as has $\rho (r)$, at least implicitly, if we assume the density in the arm varies radially as the density of the galactic disk does. $A$ and  $\Omega_p $ are then parameters that, together with initial conditions $r(0)$ and $r'(0)$, can be used to fit the arm.  That is, we can numerically solve this equation, adjusting these values until we have a reasonable fit to images.  We have done so for a couple well known examples, Messier 101 (a grand design trailing arm spiral -- figure 5) and NGC 5422 (which is believed to have leading outer arms -- figure 6).  For Messier 101 we assumed a flat rotation curve (arbitrarily chosen as $V(r)=2$).  NGC 5422 has an increasing rotation curve -- we fit the available data points with a cubic, then used that for the $V(r)$.  

Numerical solutions to this equation also reveal the connections between rotation curves, pattern speeds, and arm shape in this model.  In figure 7 we have used a flat rotation curve, fixed $A$, $r(0)$, and $r'(0)$, and varied the pattern speed  $\Omega_p $.  We can see that if we knew each of the fixed values, we could, in principle, determine the pattern speed from the shape of the arm. 

In this analysis there is nothing surprising about leading arm structures -- they are simply different values of the parameters and different rotation curves.

A significant percentage of spiral galaxies are barred spirals -- they have a bar in the center, and the spiral arms originate at the end of the bar.  These bars seem to be reasonably modeled as ellipsoids.  We observe that if think of the mass in an equidistributed ellipsoid as projected normally onto a line along the major axis, this gives a quadratic distribution, the distribution that can be shown, via the above equation of motion for a filament, to be the one required for rigid rotation.  Moreover the rotation curves for matter near the center of the disk often seem to rise roughly linearly, as one would expect for rigid rotation. So we could perhaps model the entire structure of  barred spirals with the filament equation of motion. 

It is well known that there does not appear to be enough matter in a typical disk galaxy to hold it together, given the rates at which they appear to spin.  This has led to the theory of dark matter, additional mass which cannot be 'seen', but which provides gravitational stability.  The distribution of this mass in a typical galaxy is a matter of study.  While we do not directly address dark matter here, it may be that the model has some implications in the discussion, in that we could ask which distributions are consistent with the data (rotation curves and spiral structure) and the model.

\bigskip
\centerline{References}
\medskip

1. Binney, J. and Tremaine, S. 1987, Galactic Dynamics (Princeton, NJ: Princeton Univ. Press)
\medskip

2. Bertin G., Lin C. C., 1996, Spiral Structure in Galaxies: A Density Wave Theory. MIT Press, Cambridge, MA 
\medskip

3. Toomre A., 1964, ApJ, 139, 1217
\medskip

4. Sharon E. Meidt et al. 2009 ApJ 702 277 
\medskip

5. K. Foyle et al. 2011 ApJ 735 101 
\medskip

6. J. A. Sellwood 2012 ApJ 751 44
\medskip

7. Ferreras, I., Cropper, M., Kawata, D., Page, M. and Hoversten, E. A. (2012),  MNRAS, 424: 1636–1646

8.  H. Salo et al. 2010 ApJ 715 L56 
\medskip

9. C. L. Dobbs, C. Theis, J. E. Pringle, M. R. Bate MNRAS, vol. 403, no. 2, pp. 625-645, 2010

10. Buck, G (to appear)
\medskip

11. Buck, G Nature 395, 51-53
\medskip

12. Block D. L., Puerari I., Frogel J. A., Eskridge P. B., Stockton A., Fuchs B., 1999, ApSS, 269, 5 
\medskip

13. Seigar M. S., Block D. L., Puerari I., Chorney N. E., James P. A., 2005, MNRAS, 359, 1065 
\medskip

14. Seigar, M.S., 2005 MNRAS; Letters. 361 (1), pp. L20-L24. 
\medskip

15. Byrd, G. G. et al. 1989, Celest. Mech., 45, 31
\medskip

16. Byrd, G. G. , Freeman, T. and Buta, R. 2006, AJ, 131, 1377
\medskip

17. Ronald J. Buta et al. 2003 AJ,125 634 doi:10.1086/345821
\medskip

\bigskip

\centerline{Figure Legends}

\bigskip

Figure 1
\medskip

There are three curves, A, B and C.  B is the future image of A under a flat rotation curve.  We can see that the differential rotation has stretched the curve and increased the wrapping around the center.  The arrows originating on B are normal vectors proportional to the curvature, as given by the self -gravitation of the curve.  The curve C is the image of curve B under this force.  C is also the future image of the curve A under rigid (no shear) rotation.  

\bigskip

Figure 2
\medskip

Schematics of arm 'resonance'.  Horizontal lines represent circular motion.  The more vertical slanted lines represent arms.  The first diagram is the trailing arm situation.  Line A is sent to line B by differential rotation.  Line B is sent to line C by normal motion (decreasing with distance from the center as curvature does).  The combination sends line A to line C, rigid translation.  The second diagram is the leading arm situation.  Line A is sent to line B by differential rotation, though in this case outside rotation is greater than inside rotation.  Line B is sent to line C by normal motion (decreasing with distance from the center as curvature does).  The combination sends line A to line C, rigid translation.  In this case the arm pattern rotational speed is greater than the disk rotation speeds.  The third diagram shows that greater shear (consider the distance between lines A and B -- it is much greater at the bottom than it is at the top) can be compensated for by greater pitch angle of the arm.
\bigskip

Figure 3

\medskip

Time spent in/near an arm depends on the curvature of the arm. A.  Masses sent toward arms of different curvature ($\kappa = 0,1,2$) with same speed and distance from arm.   The force is calculated as an inverse square integral over the filamentary arm, with an $\epsilon$ cutoff near the point of intersection of the arm and the trajectory of the mass (the trajectory is along the horizontal axis).  B.  The solutions are approximated by simple euler approximation. Vertical axis is position, with $0$ the intersection point with the arm, horizontal axis is time. Curve with fastest escape is $\kappa =0$, slower is $\kappa =1$, $\kappa =2$ might be said to not have escaped (yet). 

\bigskip

Figure 4

\medskip

Snapshots of numerical many-body simulations.  The masses are restricted to their original radius and so move in circular orbits.  In the first shots they are driven only by the differential rotation given by a flat rotation curve, and we clearly see the decoherence of the original filamentary structure by the winding.  In the next shots the only change in the system is that we have added a small change in the rotational velocities by a discretized version of the curvature between nearest neighbors.  The persistent spiral structure is evident.

\bigskip

\bigskip

Figure 5

\medskip

Modelling a spiral arm of Messier 101. We have used a flat rotation curve.

\bigskip

Figure 6

\medskip

A. Modelling one of the exterior spiral arms (thought to be leading) of NGC 5422. B. For the rotation curve we have used a cubic approximation to data points taken from reference 17 (the cubic approximation is pictured).

\bigskip

Figure 7

\medskip

Gallery of spiral arms given by numerical solution to arm equation.  Here we have fixed flat rotation curve and have varied pattern speed.  Of course other combinations are possible.

\bigskip

\bigskip

\bigskip

Department of Mathematics

Saint Anselm College

gbuck@anselm.edu

greg@gregorybuck.com

￼
￼
￼
￼
￼
￼
￼
\includegraphics{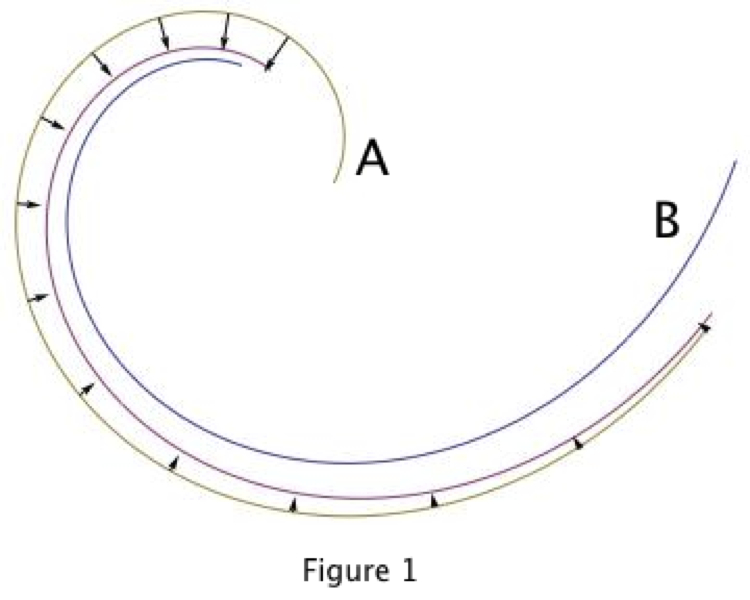}
\newpage
\includegraphics{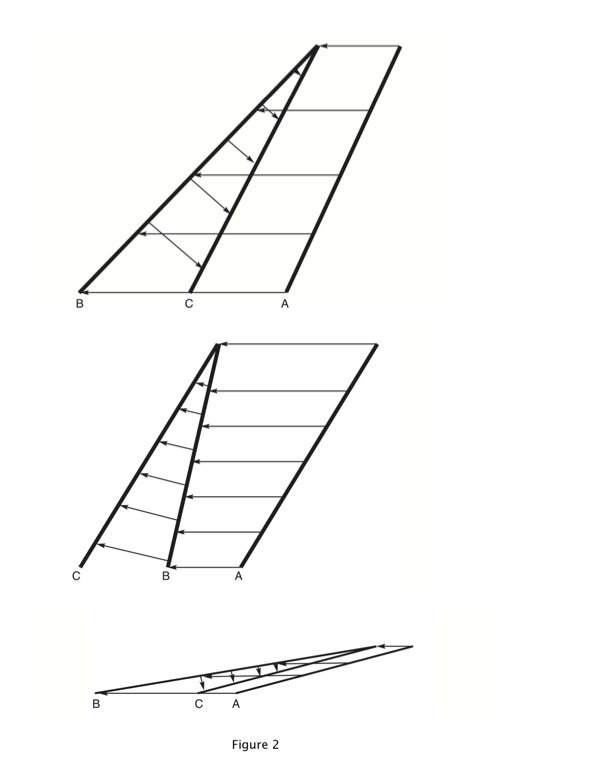}
\newpage
\includegraphics{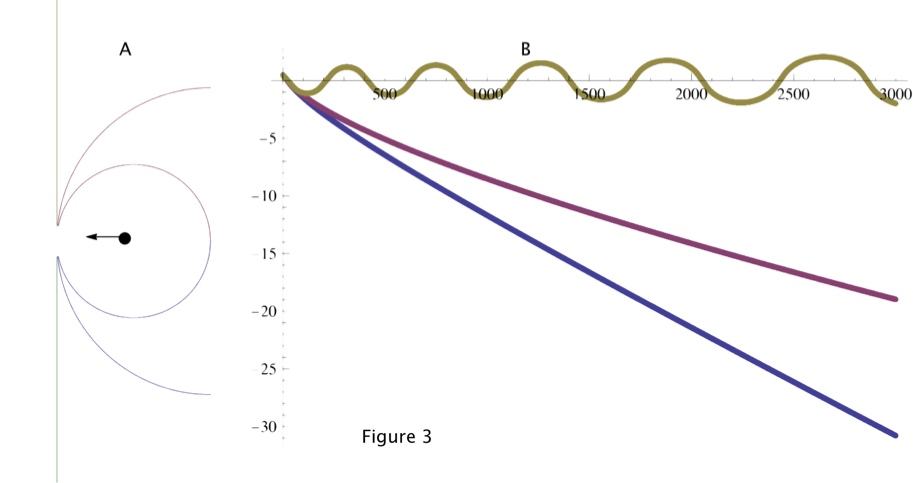}
\newpage
\includegraphics{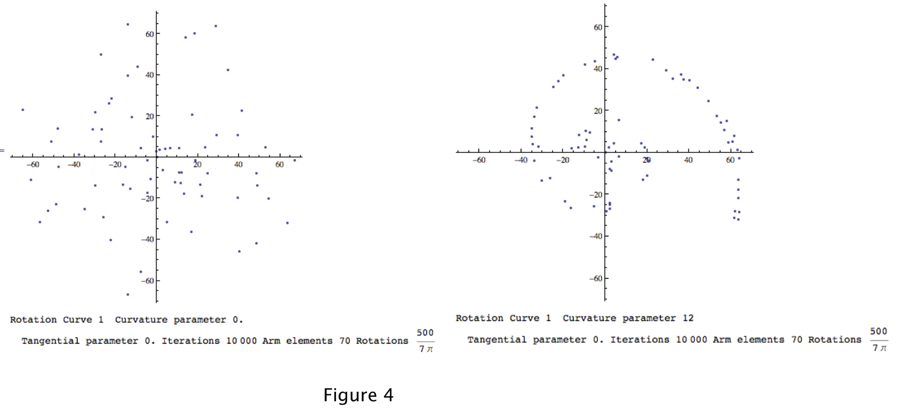}
\newpage
\includegraphics{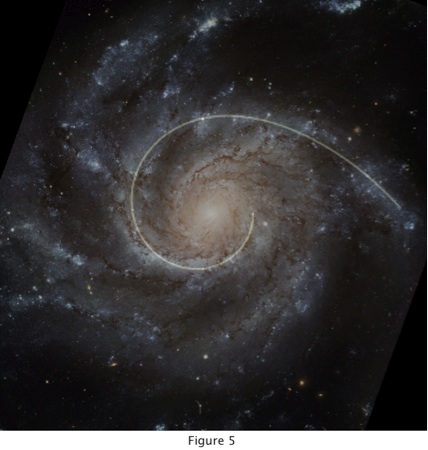}
\newpage
\includegraphics{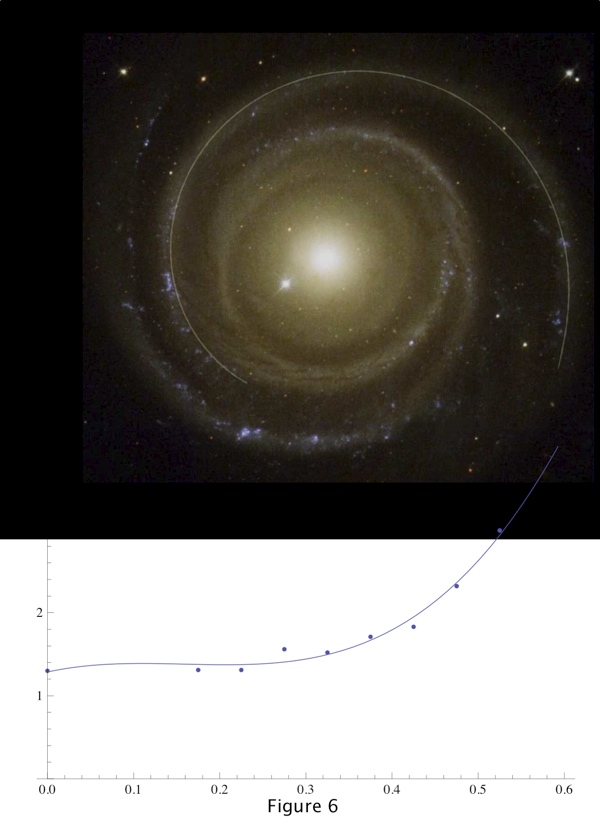}
\newpage
\includegraphics[height=240mm]{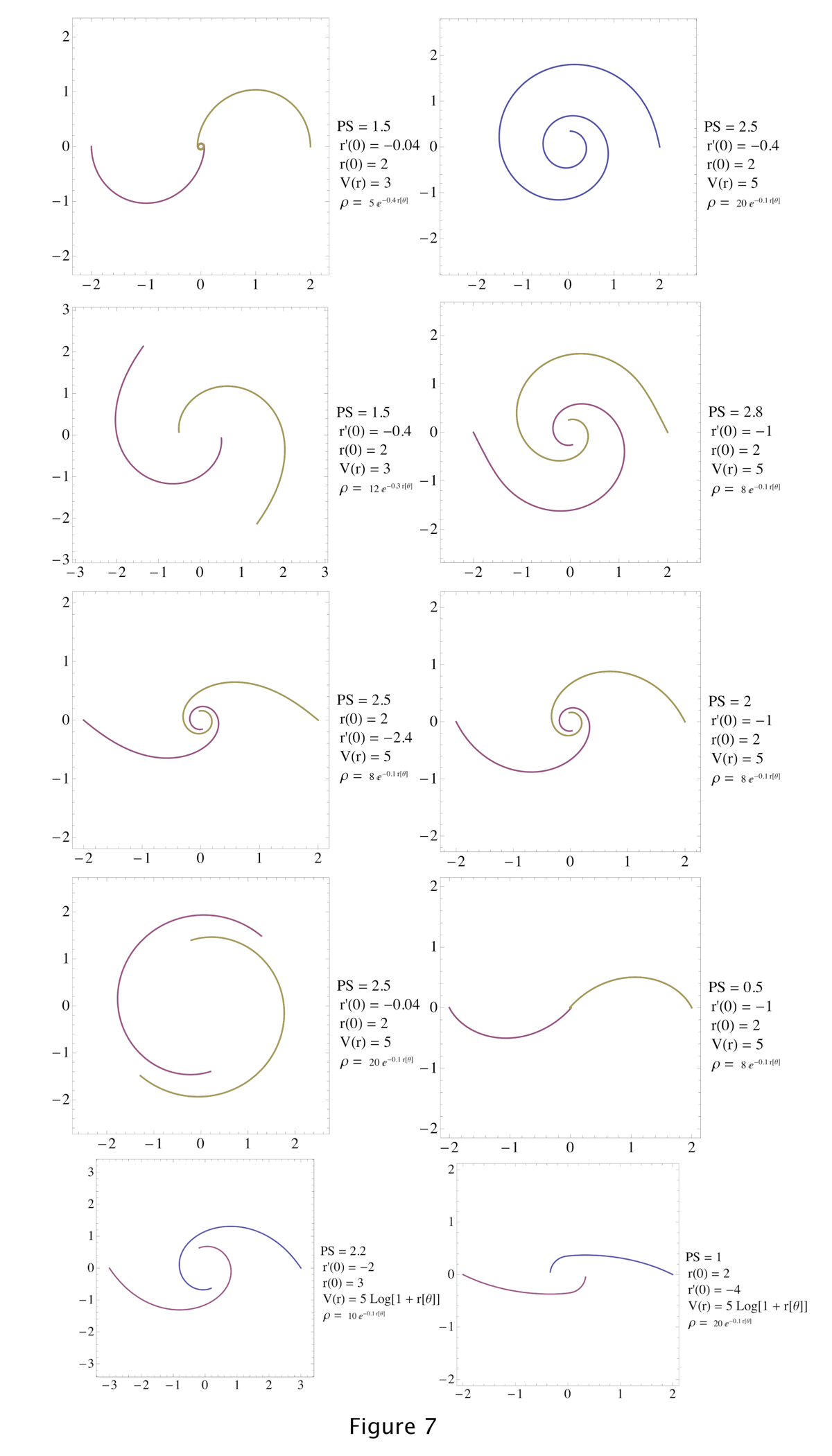}
￼
￼
￼
￼
￼
￼
￼

\end{document}